\documentclass[12pt]{article}
\usepackage[dvipdfmx]{graphicx}
\usepackage{color,amssymb,amsmath,ascmac}
\usepackage{slashed}
\usepackage[a4paper, total={16.5cm,22cm}]{geometry}
\usepackage{hyperref}

\begin{document}

\setlength{\textwidth}{16.5cm}
\setlength{\textheight}{21.5cm}
\setlength{\oddsidemargin}{0cm}
\setlength{\evensidemargin}{0cm}
\setlength{\topmargin}{0cm}
\setlength{\footskip}{1cm}

\newcommand{\lrf}[2]{ \left(\frac{#1}{#2}\right)}
\newcommand{\lrfp}[3]{ \left(\frac{#1}{#2} \right)^{#3}}
\newcommand{\vev}[1]{\left\langle #1\right\rangle}

\newcommand{\TeV}{\text{TeV}}
\newcommand{\GeV}{\text{GeV}}
\newcommand{\MeV}{\text{MeV}}
\newcommand{\keV}{\text{keV}}
\newcommand{\eV}{\text{eV}}

\begin{titlepage}
\begin{flushright}
UT-16-19\\
IPMU-16-0063
\end{flushright}
\vskip 2cm
\begin{center}
  {\Large \bf Probing the origin of 750 GeV diphoton excess \\
    with the precision measurements at the ILC\\ 
  }
\vskip 1.5cm
{
Kyu Jung Bae$^{(a)}$,
Koichi Hamaguchi$^{(a,b)}$, 
Takeo Moroi$^{(a,b)}$
and
Keisuke Yanagi$^{(a)}$
}
\vskip 0.9cm
{\it $^{(a)}$ Department of Physics, University of Tokyo, Bunkyo-ku, Tokyo 113--0033, Japan \vspace{0.2cm}
\par
$^{(b)}$ Kavli Institute for the Physics and Mathematics of the Universe (Kavli IPMU), \\
University of Tokyo, Kashiwa 277--8583, Japan
}
\vskip 2.5cm
\abstract{

  The recently reported diphoton excess at the LHC may imply the
  existence of a new resonance with a mass of about 750 GeV which
  couples to photons via loops of new charged particles.  In this
  letter, we study the possibility to test such models at the ILC,
  paying attention to the new charged particles responsible for the
  diphoton decay of the resonance.  We show that they affect 
   the scattering processes $e^+e^- \to f\bar{f}$ 
   (with $f$ denoting Standard Model fermions)
   at the ILC, which makes it possible to indirectly probe the new charged
  particles even if they are out of the kinematical reach. We also
  show that the discriminations of the diphoton models may be possible
  based on  a study of the angular distributions of $f\bar{f}$.
}

\end{center}
\end{titlepage}

\section{Introduction}

The ATLAS and CMS collaborations reported an excess of diphoton
events, which suggests an existence of a new resonance with a mass of
around 750 GeV~\cite{ATLAS-CONF-2015-081,CMS:2015dxe}.  One of the
natural explanations of the excess is by the production and decay of a
(pseudo-) scalar particle $S$ with a mass of $\sim 750$ GeV, $pp\to
S\to \gamma\gamma$, assuming that the excess is not due to a
statistical fluctuation.  If the excess is confirmed with higher
statistics in the near future, the high-priority task is to understand
the nature of the diphoton resonance and physics behind it.  One
important question is the origin of the interaction of the scalar
particle $S$ with photon (and other SM gauge bosons).

In most of the scenarios, the particle $S$ is not the only particle at
the TeV scale, but there are also new charged particles with masses of
${\cal O}(\TeV)$ or smaller which are responsible for inducing the
coupling between the $S$ and photons via loop effects.
Those new charged particles are important targets of future collider
experiments.  Although we hope to find them at the LHC run 2, the mass
reach via the direct searches is strongly model dependent.  In
particular, if non-colored new particles are responsible for the
coupling between the scalar $S$ and photons, their direct production
cross section at the LHC is suppressed and they may not be easily
detected at the LHC.

Using the fact that the charged particles contribute to the vacuum
polarization of the SM gauge bosons, we may indirectly probe the new
charged particles.  In particular, with high statistics and clean
environment, the future International $e^+e^-$ Linear Collider (ILC)
\cite{Behnke:2013xla,ILCpapers} can provide very accurate
information about the vacuum polarization through detailed studies of
the scattering and pair-production processes of SM fermions,
$e^+e^-\to f\bar{f}$~\cite{Harigaya:2015yaa}. Notably,
even if the new charged particles are kinematically inaccessible,
their contribution to the vacuum polarization of the SM gauge bosons
may be large enough to be probed by the precise measurements at the
ILC.  Such a study gives very important information to reveal the
nature of the diphoton resonance.\footnote
{For the possibility of directly studying the diphoton resonance at
  the ILC, see \cite{diphoton_ILC_papers}.}

In this letter, we investigate the possibility of the indirect probe
of the new particles at the ILC, which is complementary to the direct
search at the LHC.  A crucial point here is that the diphoton excess
requires a large multiplicity and/or a large charge of the new
particles in the loop, especially when their mass is large, and such a
large multiplicity and/or a large charge enhance the ILC signal.  We
apply the analysis of \cite{Harigaya:2015yaa} to diphoton models, and
show that a large parameter region the models can be covered
by using the ILC precision measurement.  We also study the possibility
to probe the gauge quantum numbers of the new particles by using the
angular distribution of the final states of the scatting processes.

The rest of this letter is organized as follows.  In
Sec.~\ref{sec:setup} we show our setup and introduce simplified models
for the diphoton excess.  Our main analysis is presented in
Sec.~\ref{sec:main}, where the ILC reach for the diphoton models are
estimated.  In Sec.~\ref{sec:SU2vsU1}, we study the possibility to
probe the ${\rm SU}(2)\times {\rm U}(1)_Y$ representation of the new
charged particles.  Sec.~\ref{sec:summary} is devoted to summary and
discussion.

\section{Setup}
\label{sec:setup}

We assume that the coupling between the 750 GeV (psuedo-) scalar 
$S$ and the photon is
induced by a diagram with new charged fermions running in the loop.
For simplicity, we assume that there are $N$ copies of fermions
$\psi_i$, all of which transform as $n$-plet under SU(2), have a
U(1)$_Y$ charge $Y$, a common mass $m$ and a common Yukawa
coupling $y$ to the scalar $S$:
\begin{align}
  {\cal L}_{\psi} &= \sum_i \bar{\psi}_i(i \slashed{D}-m)\psi_i - 
  i \sum_i y S\bar{\psi}_i \gamma_5 \psi_i\,,
\end{align}
where we assume that $S$ is a pseudoscalar.  
In the case of scalar $S$, the
second term is replaced with $\sum_i y S\bar{\psi}_i
\psi_i$.  The following discussion does not depend on whether or not
the fermions $\psi_i$ have an SU(3) charge. Hereafter, the
multiplicity $N$ is understood to include the color factor.

In our analysis, we further assume that the $S$ mainly decays into
gluon pairs:
\begin{align}
  \Gamma(S;\text{total})\simeq \Gamma(S\to gg)\gg \Gamma(S\to \gamma\gamma)\,.
  \label{gg>gammagamma}
\end{align}
Then, the diphoton signal rate is given by
\begin{align}
\sigma(pp\to S\to \gamma\gamma)
&\simeq
\frac{C_{gg}}{s\, m_S}\Gamma(S\to \gamma\gamma),
\end{align}
where $\sqrt{s}=13\ {\rm TeV}$ and $C_{gg}=(\pi^2/8)\int^1_0 dx_1
\int^1_0 dx_2$ $\delta(x_1x_2-m_S^2/s)g(x_1)g(x_2)$ with $g(x)$ being
the gluon parton distribution function.  In our numerical calculation,
we use the MSTW2008 NLO set~\cite{Martin:2009iq} evaluated at the
scale $\mu=m_S$, which gives $C_{gg}\simeq 2.1\times 10^3$.  Thus, the
diphoton signal rate is determined by the partial decay rate
$\Gamma(S\to \gamma\gamma)$, which is given by
\begin{align}
\Gamma(S\to \gamma\gamma)
&\simeq
\frac{\alpha^2}{256\pi^3}
m_S^3
\left[
 \frac{y}{m} \text{tr}Q^2
 L\lrf{m_S^2}{4m^2}
\right]^2 ,
\end{align}
where 
$\alpha$ is the fine structure constant.
The loop function is given by
\begin{align}
L(\tau) &=
\begin{cases}
2\tau^{-2}\left(\tau+(\tau-1)\arcsin^2\sqrt{\tau}\right) & \text{for scalar }S\,,
\\
2\tau^{-1} \arcsin^2\sqrt{\tau} & \text{for pseudo-scalar }S\,,
\label{eq:loop-function}
\end{cases}
\end{align}
and the trace of electric charge $\text{tr}Q^2$ is defined as
\begin{align}
\text{tr}Q^2 = N\left[ \frac{n(n-1)(n+1)}{12} +nY^2\right]\,.
\label{trQ^2}
\end{align}
Note that the multiplicity $N$ includes the possible color factor.  In order to
realize $\sigma(pp\to S\to \gamma\gamma)=3$ -- $10\ {\rm fb}$, the
partial decay width is required to be $\Gamma(S\to
\gamma\gamma)=0.45$ -- $1.5\ {\rm MeV}$, assuming Eq.\
\eqref{gg>gammagamma}.

Before discussing the ILC signals in the next section, 
let us exemplify some simple models
which may be difficult to probe by the direct searches at the LHC
but can be tested by the ILC indirect measurement studied in this work.
As an example, suppose that $\psi_i$ have the same quantum number
as the SM right-handed leptons, i.e., singlet under 
 ${\rm SU}(3)\times {\rm SU}(2)$ and $Y=1$.
For instance, $(N,y,m)\simeq (7,0.3,400~\GeV)$ 
or $(5,1,650~\GeV)$ can lead to $\Gamma(S\to \gamma\gamma)\simeq 1.0~\MeV$.
The direct search at the LHC strongly depends on their decay modes.
Let us assume that they are mainly coupled with the left-handed tau leptons,
via a small Yukawa coupling with the SM Higgs.
The prospects for excluding or discovering such a vector-like lepton at the LHC
are studied in Ref.~\cite{Kumar:2015tna}, which shows that, 
even in the optimistic scenario that the background is known exactly, 
it would take 1000 fb$^{-1}$ to exclude up to $m = 200~\GeV$.
Although the multiplicity $N>1$ increases the number of signal events, 
we expect that heavier mass region
is very difficult to probe even with higher integrated luminosity.
Similarly, we can also consider the case that $\psi_i$ have the same quantum number
as the SM left-handed leptons, mainly coupled to the right-handed tau leptons.
Ref.~\cite{Kumar:2015tna} showed that 95\% C.L. exclusion up to $m\simeq 440~\GeV$ is possible
at 13 TeV LHC with 100 fb$^{-1}$, but again it will be challenging to reach 
heavier mass region such as $m\simeq 600~\GeV$.

It is also easy to satisfy the assumption in Eq.~\eqref{gg>gammagamma}. 
If the charged particles $\psi_i$ are
non-colored and/or its contribution to $\Gamma(S\to gg)$ is not
sufficient, additional colored particles which couples to $S$ may be
introduced.  For example, one can consider that 
the coupling between $S$ and gluons is induced by a
Majorana fermion, $\widetilde{g}$, which
transforms as the adjoint representation of SU(3), like the gluino in
SUSY models.  Then, the decay rate of the psuedo-scalar $S$ into gluons is given by
$\Gamma(S\to gg) \simeq 3~\MeV \times N_{\widetilde{g}}^2
 y_{\widetilde{g}}^2 (m_{\widetilde{g}}/3~\TeV)^{-2}$,
where $m_{\widetilde{g}}$, $N_{\widetilde{g}}$, and $y_{\widetilde{g}}$
are the mass, the multiplicity, and the Yukawa coupling to $S$, respectively.
Thus, the condition $\Gamma(S\to
gg)\gg\Gamma(S\to\gamma\gamma)$ can easily be satisfied, e.g.,
by $N_{\widetilde{g}}=2$, $y_{\widetilde{g}}\simeq 1$, and $m_{\widetilde{g}}\simeq 3~\TeV$.
Such a heavy particle is
difficult to probe at the LHC.  The fermion $\widetilde{g}$ can
decay into e.g., three quarks via an exchange of a heavy colored
scalar (like the squark in SUSY models), and can easily satisfy the
cosmological constraints.

\section{Indirect signals at ILC}
\label{sec:main}

We consider the case that the masses of the new charged particles are
larger than the beam energy and kinematically inaccessible, i.e.,
$\sqrt{s}<2m$.  Even in such a case, the new charged particles
affects the observables at the ILC through radiative corrections.  In
particular, we pay attention to the contributions to the vacuum
polarizations of standard model gauge bosons.

Because we are interested in the case where the interactions of the
new charged particles with the Higgs fields are negligible for the ILC
processes, we only have to consider the vacuum polarizations of SU(2)
and U(1)$_Y$ gauge bosons.  With the set up given in the previous
section, the contributions of the new particles to the vacuum
polarizations are given by
\begin{align}
  \delta \Pi_{VV} (q^2,m^2) \equiv \frac{1}{2} g_V^2 C_{VV} I (q^2/m^2),
\end{align}
with $V=W$ (for SU(2)) and $B$ (for U(1)$_Y$), where $g_V$ is the
gauge coupling constant for SU(2) or U(1)$_Y$, $q$ is the four
momentum of the gauge bosons, $m$ is the mass of the new charged
fermions,
\begin{align}
  \label{eq:Pi}
  I (x) \equiv
  \frac{1}{16\pi^2}
    \displaystyle{\int^1_0 dy~y(1-y)\mathrm{ln}(1-y(1-y)x)}
\end{align}
and the coefficients are given by
\begin{align}
  C_{WW} &= \frac{4}{3} N n(n-1)(n+1)\,,
  \label{C_WW}  
  \\
  C_{BB} &= 16nNY^2\,.
  \label{C_BB}  
\end{align}
If the particle in the loop is a Majorana fermion with a real representation, such as $({\bf 1},{\bf 3},0)$,
an additional factor of $1/2$ is necessary for $C_{WW}$.\footnote{In the case of scalar loop, there is an additional factor of $1/8$ for both $C_{WW}$ and $C_{BB}$ and the function $I(x)$ in Eq.~\eqref{eq:Pi} becomes
$I(x)=(1/16\pi^2)\int^1_0 dy~(1-2y)^2 \ln(1-y(1-y)x)$.
See the comments at the end of this section.
\label{fn:scalar} }
 For the convenience of the
following discussion, we define the ratio:
\begin{align}
  R_{21} \equiv  C_{WW}/C_{BB} = \frac{n^2-1}{12Y^2}.
\end{align}
Notice that $R_{21}$ corresponds to the ratio of the SU(2) and
U(1)$_Y$ contributions to ${\rm tr}Q^2$ (see Eq.\ \eqref{trQ^2}).

These new contributions to the vacuum polarization affect the
scattering processes at the ILC.  We investigate the corrections to
the SM process, $e^+e^-\to  f\bar{f}$, taking into
account the new charged particles running in the vacuum polarization
loop. In our analysis, we concentrate on the final states of $e^+e^-$
and $\mu^+\mu^-$.

As in the analysis of Ref.~\cite{Harigaya:2015yaa}, we define bins to
use the information about the angular distribution of the final state
particles of the process mentioned above.  The bins are defined by ten
uniform intervals for the scattering angle $\cos\theta$, $-1\le
\cos\theta\le 1$ for the $\mu^+\mu^-$ final state and $-0.99\le
\cos\theta\le 0.99$ for the $e^+e^-$ final state.  Then, we study
the expected sensitivity of the ILC by calculating the following quantity:
\begin{align}
  \Delta \chi^2 & = \sum_{i:\,{\rm bins}}
  \frac{(N_i^{\rm SM+\psi}-N_i^{\rm SM})^2}{N_i^{\rm SM}+(\epsilon N_i^{\rm SM})^2},
\end{align}
where $\epsilon$ is the systematic uncertainty, and $N_i^{\rm SM}$ and
$N_i^{\rm SM+\psi}$ are the expected numbers of events in $i$-th bin
based on the SM and the model with the new particles,
respectively. $N_i^{\rm SM}$ and $N_i^{\rm SM+\psi}$ are calculated
with the amplitudes ${\cal M}^{\rm SM}$ and ${\cal M}^{\rm
  SM+\psi}\equiv{\cal M}^{\rm SM}+{\cal M}^{\psi}$, respectively; the
explicit formulae of the amplitudes are given by
\begin{align}
  \label{eq:M_mu}
  {\cal M}^{\rm SM,\psi}
  ( e^-_{h} e^+_{\bar{h}} \rightarrow \mu^-_{h'} \mu^+_{\bar{h}'} )
  =
  \sum_{V,V'=\gamma,Z} C_{e_hV}C_{\mu_{h'}V'}
  D^{\rm SM, \psi}_{VV'} (s)
  [\bar{u}_{h'} \gamma^{\mu} v_{\bar{h}'}] 
  [\bar{v}_{\bar{h}} \gamma_{\mu} u_{h}] ,
\end{align}
and
\begin{align}
  \label{eq:M_e}
  {\cal M}^{\rm SM, \psi} 
  \left( e^-_h e^+_{\bar{h}} \rightarrow e^-_{h'} e^+_{\bar{h}'} \right) =&
  \sum_{V,V'=\gamma,Z} C_{e_hV}C_{e_{h'}V'}
  D^{\rm SM, \psi}_{VV'}(s) 
  [\bar{u}_{h'} \gamma^{\mu} v_{\bar{h}'}] 
  [\bar{v}_{\bar{h}} \gamma_{\mu} u_{h}] 
  \nonumber \\ &
  - \sum_{V,V'=\gamma,Z} C_{e_hV}C_{e_{h'}V'} 
  D^{\rm SM, \psi}_{VV'}(t) 
  [\bar{u}_{h'} \gamma^{\mu} u_h]
  [\bar{v}_{\bar{h}} \gamma_{\mu} v_{\bar{h}'}],
\end{align}
where $u_{h}$, $\bar{v}_{\bar{h}}$, $v_{\bar{h}'}$, and $\bar{u}_{h'}$
are spinors for initial and final state particles (with $h^{(\prime)}$
and $\bar{h}^{(\prime)}$ being the helicities), $t\equiv (p-p')^2$
(with $p$ and $p'$ denoting the momenta of initial- and final-state
leptons, respectively), $C_{f_hV}$ are coupling constants of incoming
and outgoing fermions with gauge bosons, defined as
\begin{align}
  &
  C_{e_L Z} = C_{\mu_L Z} = g_Z( -1/2 + \sin^2 \theta_W ),
  \\
  &
  C_{e_R Z} = C_{\mu_R Z} = g_Z \sin^2 \theta_W,
  \\
  &
  C_{e_L \gamma } = C_{e_R \gamma } = C_{\mu_L \gamma } = C_{\mu_R \gamma } = -e,
\end{align}
with $e$ being the electric charge, $\theta_W$ the Weinberg angle, and
$g_Z = e/\sin\theta_W\cos\theta_W$.  In addition,\footnote
{For simplicity, we use the leading order SM amplitude in our
  analysis.  We have checked our LO calculation reproduces the results
  of Ref.~\cite{Harigaya:2015yaa}, which is based on NLO formulae for
  $D^{\rm SM}_{VV'}$, within a few \% difference in the mass reach for
  the new fermions.}
\begin{align}
  \label{eq:Mtilde_mu}
  D^{\rm SM}_{VV'} (q^2) &= \frac{\delta_{VV'}}{q^2 - m_V^2}, \\
  D^\psi_{VV'} (q^2) &=
  \frac{q^2}{(q^2-m_V^2)(q^2-m_{V'}^2)} \delta\Pi_{VV'} (q^2, m),  
\end{align}
where
\begin{align}
  \delta\Pi_{\gamma\gamma} (q^2, m) & = 
  \delta\Pi_{WW} (q^2, m)  \sin^2 \theta_W
  + \delta\Pi_{BB} (q^2, m)  \cos^2 \theta_W,
  \\ 
  \delta\Pi_{ZZ} (q^2, m) & = 
  \delta\Pi_{WW} (q^2, m)  \cos^2 \theta_W
  + \delta\Pi_{BB} (q^2, m)  \sin^2 \theta_W,
  \\ 
  \delta\Pi_{\gamma Z} (q^2, m) & = 
  \left[ \delta\Pi_{WW} (q^2, m) - \delta\Pi_{BB} (q^2, m) \right]
  \sin\theta_W \cos\theta_W.
\end{align}
We comment here that $\delta\Pi_{VV}$ becomes more enhanced with 
larger charge and larger multiplicity of the new particles, which are favored
to explain the diphoton excess at the LHC.  As we will see below, the mass
reach for the new particles becomes better in such a parameter space.

We evaluate $\Delta \chi^2$ and determine the mass reach for the new
charged particles at the ILC.  The center-of-mass energy is taken to
be $\sqrt{s} = 500 \mathrm{GeV}$ and $\sqrt{s} = 1 \mathrm{TeV}$.  The
beam polarization of incoming electron is taken to be $-80\%$, while
that of positron is chosen as $+30\%$ \cite{Behnke:2013xla}.  The
integrated luminosity is taken to be $1$ -- $3 {\rm ab}^{-1}$.

\begin{figure}
\centering
    \includegraphics[width=16cm]{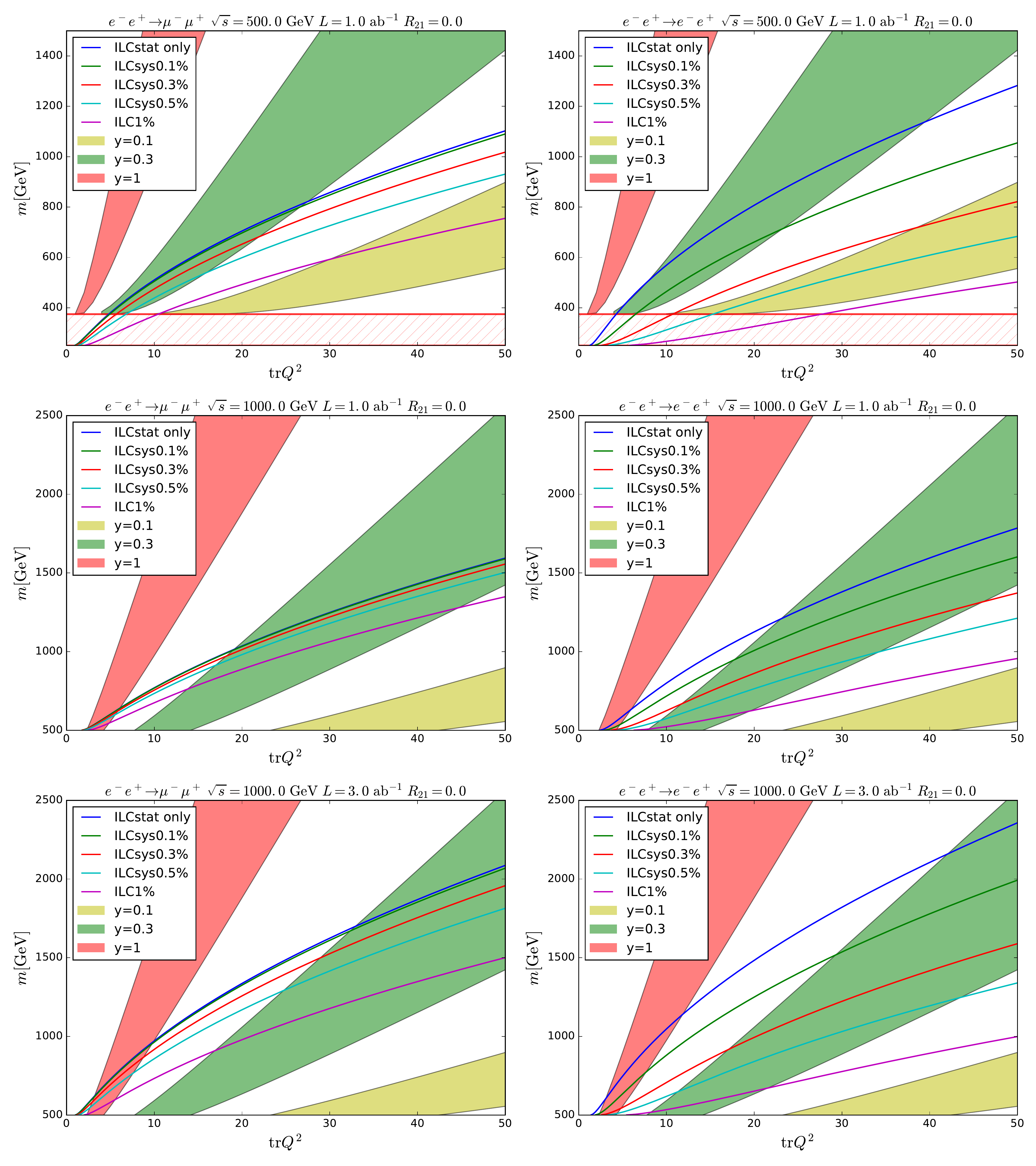}
    \caption{Mass reach for $R_{21}=0$. The ILC beam energy is
  $\sqrt{s}=500\mathrm{GeV}$ ($1\mathrm{TeV}$) for the upper figures
  (the middle and lower figures), while the integrated luminosity is
  $1\ {\rm ab}^{-1}$ ($3\ {\rm ab}^{-1}$) for the upper and middle
  figures (lower figures).  Three figures on the left are ILC bounds
  from the $\mu^+\mu^-$ final state, and right figures are bounds from
  the $e^+e^-$ final state.  The region below each solid line can be
  probed by the ILC with each systematic uncertainty.  The yellow,
  green, and red bands show the region in which $\sigma(pp\to S\to
  \gamma\gamma)=3$ -- $10$ fb is realized with $y=0.1$, $0.3$, and
  $1$, respectively.}
  \label{fig:R=0}
\end{figure}

\begin{figure}
\centering
    \includegraphics[width=16cm]{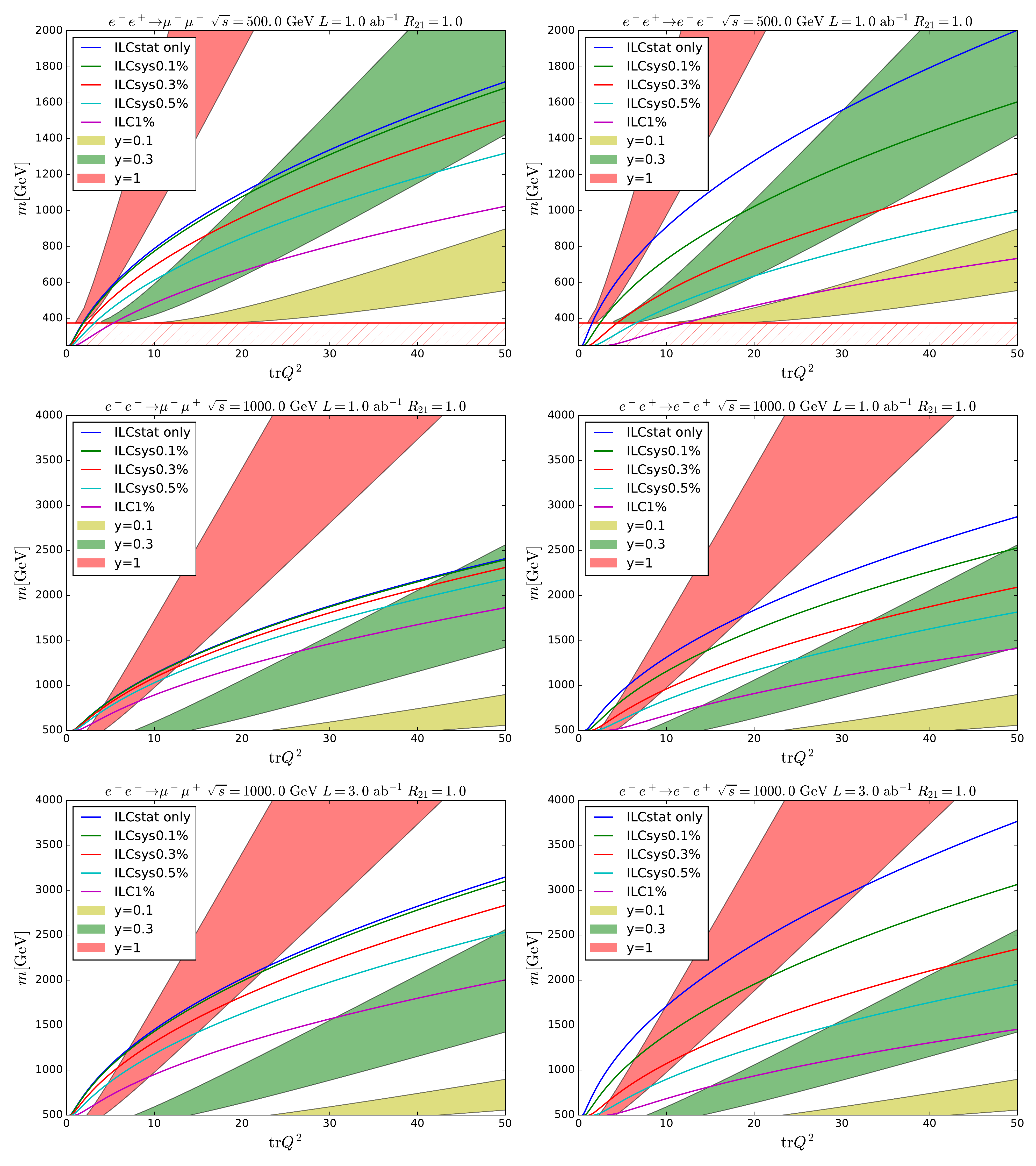}
    \caption{Same as Fig.\ \ref{fig:R=0}, except for $R_{21}=1$.}
\label{fig:R=1}
\end{figure}

\begin{figure}
\centering
    \includegraphics[width=16cm]{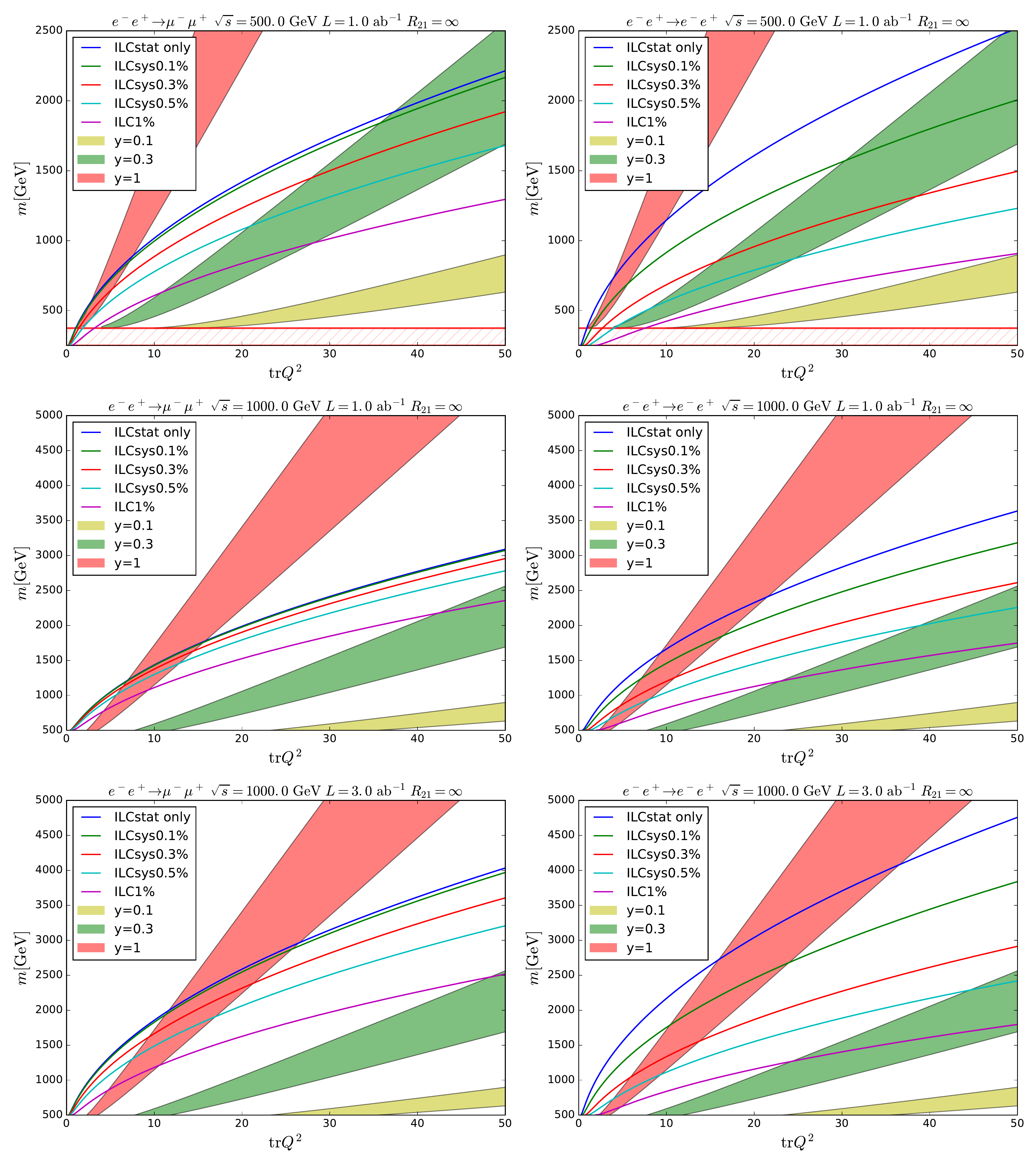}
    \caption{Same as Fig.\ \ref{fig:R=0}, except for $R_{21}=\infty$, and 
the yellow, green, and red bands showing the region of
  $\sigma(pp\to S\to \gamma\gamma)=3$ -- $7$ fb for
  $y=0.1$, $0.3$, and $1$, respectively. (See the discussion in the text.)
  }
\label{fig:R=inf}
\end{figure}

In Figs.~\ref{fig:R=0} -- \ref{fig:R=inf}, we show the contours of
$\Delta\chi^2=2.71$, which gives 95\% C.L. reach of the mass, 
on $\text{tr}Q^2$
vs.\ $m$ plane (solid lines).   Each line corresponds to the
systematic uncertainty of $\epsilon = 0, 0.1, 0.3, 0.5$, and $1\%$.
In each figure, we shaded the regions at which $\sigma(pp\to S\to
\gamma\gamma)$ becomes the relevant value to explain the diphoton
excess for the Yukawa coupling $y=0.1$, 0.3, and 1.
Here, we have assumed that the $S$ is a pseudo-scalar.\footnote{In the case of scalar $S$, 
the ILC reach does not change, while the shaded bands in Figs.~\ref{fig:R=0} -- \ref{fig:R=inf} move towards a smaller mass of the charged particle by a factor of about $2/3$ because of the difference in the loop functions \eqref{eq:loop-function}.}
As seen in the figures, the indirect probe at the ILC can cover a large parameter space 
of the diphoton models.
\begin{itemize}

\item Fig.\ \ref{fig:R=0} shows the case of $R_{21}=0$, which
  corresponds to SU(2) singlet.   For instance, 
  by measuring the cross sections of $e^+e^- \to
  \mu^+\mu^-$ and $e^+e^- \to e^+e^-$
  with $\epsilon=0.1\%$, $\sqrt{s}
  = 500\ \mathrm{GeV}$ and $L=1\ {\rm ab}^{-1}$ ($\sqrt{s} = 1\ {\rm
    TeV}$ and $L=3\ {\rm ab}^{-1}$), the ILC can probe up to $m\simeq 500\
  \mathrm{GeV}$ and $460\ \mathrm{GeV}$ ($960\ \mathrm{GeV}$ and $880\
  \mathrm{GeV}$) for $\mathrm{tr}Q^2 = 10$, respectively.

\item Fig.\ \ref{fig:R=1} shows the case of $R_{21}=1$, which
  corresponds to the case that the fermions has the same quantum
  numbers as those of the SM left-handed leptons, i.e., $({\bf 1},
  {\bf 2}, 1/2)$ for ${\rm SU}(3)\times {\rm SU}(2)\times {\rm
    U}(1)_Y$.  The mass reach is larger than the case of $R_{21} = 0$,
  because the SU(2) gauge coupling is larger than the U(1)$_Y$ gauge
  coupling and that yields larger discrepancy from SM.  
  By measuring the cross sections of $e^+e^- \to
  \mu^+\mu^-$ and $e^+e^- \to e^+e^-$ with $\epsilon=0.1\%$, $\sqrt{s}
  = 500\ \mathrm{GeV}$ and $L=1\ {\rm ab}^{-1}$ ($\sqrt{s} = 1\ {\rm
    TeV}$ and $L=3\ {\rm ab}^{-1}$), 
     the ILC can probe up to $m\simeq 780\
  \mathrm{GeV}$ and $730\ \mathrm{GeV}$ ($1430\ \mathrm{GeV}$ and
  $1390\ \mathrm{GeV}$)  for $\mathrm{tr}Q^2 = 10$, respectively.  Thus, the ILC
  will be able to reach the mass at the TeV scale if $\sqrt{s}\sim 1\
  {\rm TeV}$ is available, and hence covers a large parameter space.
\item 
Fig.\ \ref{fig:R=inf} shows the case of $R_{21}=\infty$, which
  corresponds to the fermion with $Y=0$.
    In this case, as we can see, the fermions with their masses of a few
  TeV may be probed with $\sqrt{s}\sim 1\ {\rm TeV}$, and the mass
  reach becomes the largest among the examples we consider in this
  letter.  Taking $\sqrt{s} = 500\ \mathrm{GeV}$ and $L=1\ {\rm
    ab}^{-1}$ ($\sqrt{s} = 1\ {\rm TeV}$ and $L=3\ {\rm ab}^{-1}$),
  $\epsilon=0.1\%$ and $\mathrm{tr}Q^2 = 10$,
  the ILC can probe up to $1000\ \mathrm{GeV}$ and $910\ \mathrm{GeV}$
  ($1820\ \mathrm{GeV}$ and $1750\ \mathrm{GeV}$) by measuring the cross
  sections of $e^+e^- \to \mu^+\mu^-$ and $e^+e^- \to e^+e^-$,
  respectively.
  We should note that, in the case of $Y=0$, 
  the decay of $S$ into other electroweak gauge bosons are enhanced.
  In particular, the ratio of the $Z\gamma$ to $\gamma\gamma$ decay rates become
  $\text{Br}(S\rightarrow Z\gamma)/\text{Br}(S\rightarrow
    \gamma\gamma)\simeq 6.3$.
      The 8 TeV run of the LHC
    has provided an upper bound of $\text{Br}(S\rightarrow
    Z\gamma)/\text{Br}(S\rightarrow \gamma\gamma)\lesssim 8.4 \times [\sigma
    (pp\rightarrow S\rightarrow\gamma\gamma)/5\ {\rm fb}]^{-1}$
    (see, e.g., \cite{Knapen:2015dap}).  Thus, in Fig.\ \ref{fig:R=inf}
    we show the region of $\sigma (pp\rightarrow S\rightarrow\gamma\gamma)\leq 7\
    {\rm fb}$.
\end{itemize}

Before closing this section, let us briefly comment on the possibilities to
probe other scenarios.
First, assuming $S$ is CP even,
the charged particles in the loop for the diphoton signal can be scalars.
Even in such a case, the charged scalars affect the ILC processes through
their contributions to the vacuum polarizations. (See footnote \ref{fn:scalar}.)
We checked that a large parameter space of the diphoton models is probed 
also in such a case.
Next, there is a different scenario that the 750 GeV resonance is a QCD
bound state of vector-like quarks with a mass of about 375
GeV and a hypercharge $Y=-4/3$~\cite{375withY43}.
This scenario corresponds to
the point $(\text{tr}Q^2,m)=(16/3, 375~\GeV)$ in Fig.\ \ref{fig:R=0},
 which is within the reach of the ILC with
$\sqrt{s}=500~\GeV$.

\section{Studying SU(2) and U(1)$_Y$ quantum numbers}
\label{sec:SU2vsU1}

Now we consider how well we can distinguish different models
containing new particles with different gauge quantum numbers.  For
this purpose, we use the fact that, for the process $e^+e^-\rightarrow
\bar{f}f$ (with $f\neq e^-$), the effects of the new particles (with
fixed $s$) are determined by only two parameters: $\delta\Pi_{BB} (s)$
and $\delta\Pi_{WW} (s)$.  As one can understand from Eqs.\
\eqref{C_WW} and \eqref{C_BB}, the relative size of $\delta\Pi_{BB}
(s)$ and $\delta\Pi_{WW} (s)$ is sensitive to the gauge quantum
numbers of the new particles.  Importantly, the effects of
$\delta\Pi_{BB} (s)$ and $\delta\Pi_{WW} (s)$ on the angular
distributions are different.  In the following, we discuss how well we
can distinguish models behind the diphoton excess at the LHC by using
the scattering process $e^+e^-\rightarrow \mu^+\mu^-$.

\begin{figure}[t]
  \centering
  \includegraphics[width=10cm]{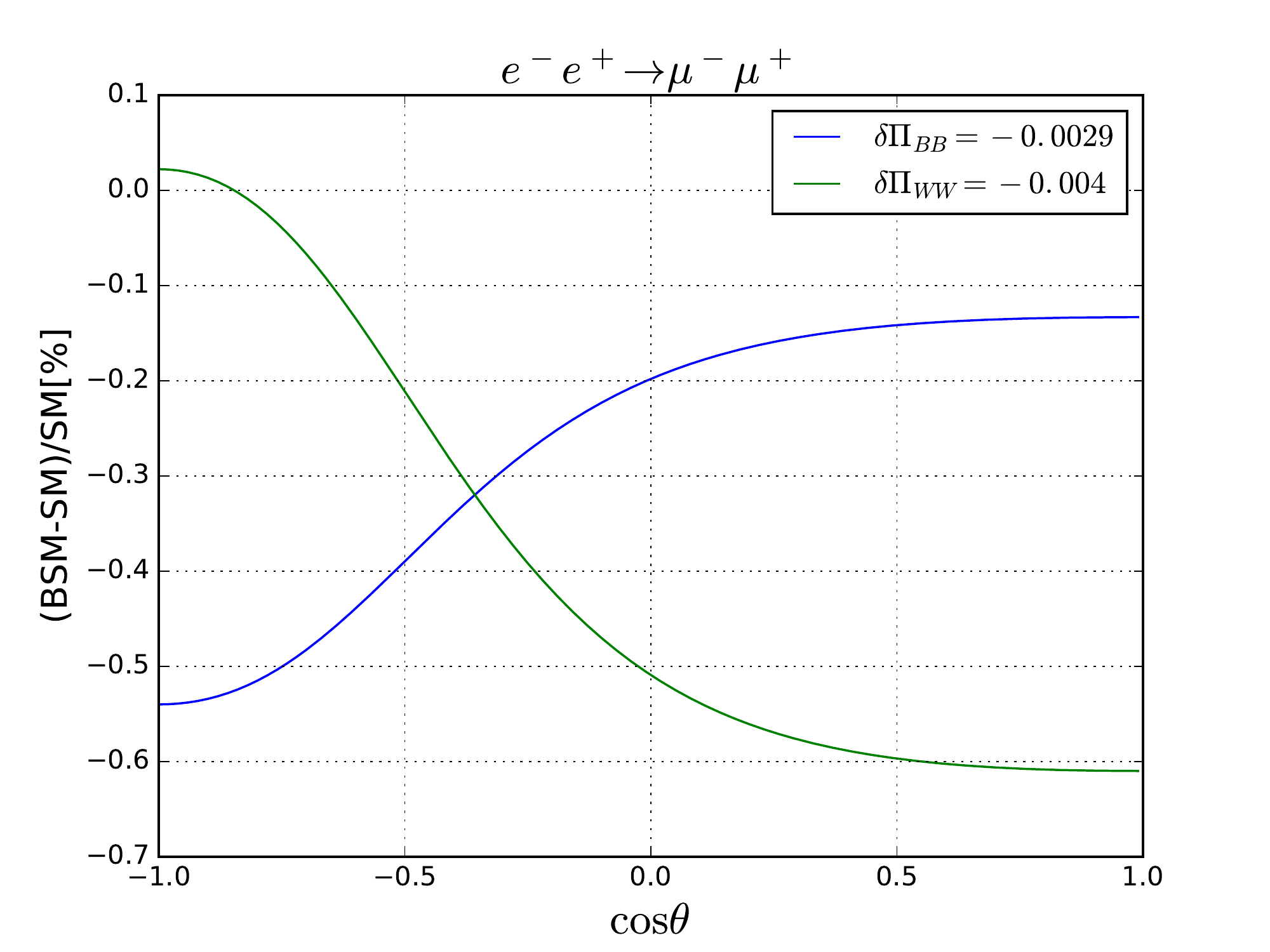}
  \caption{The deviation of the differential cross section from the
    Standard model. Blue and green lines show the cases with $(\delta\Pi_{BB},\delta\Pi_{WW})=(-0.0029,0)$ and
$(0,-0.0040)$, respectively.
 We take $\sqrt{s}=500$GeV, $P_-=-80\%$  and
    $P_+=30\%$. }
  \label{fig:ang-dep}
\end{figure}

First, for the demonstration of the angular distribution with
non-vanishing $\delta\Pi_{BB} (s)$ or $\delta\Pi_{WW} (s)$, let us
define
\begin{align}
  {\cal F}_{\mu^+\mu^-} (\cos\theta) 
  \equiv
  \frac{
    [{d \sigma^{\rm SM+\psi} (e^+e^- \to \mu^+\mu^-)}/{d\cos\theta}]
    - [{d \sigma^{\rm SM} (e^+e^- \to \mu^+\mu^-)}/{d\cos\theta}]}
  {[{d \sigma^{\rm SM} (e^+e^- \to \mu^+\mu^-)}/{d\cos\theta}]},
\end{align}
where $\sigma^{\rm SM}$ and $\sigma^{\rm SM+\psi}$ are cross sections
in the SM and in the model with the new charged particles,
respectively.  In Fig.~\ref{fig:ang-dep}, we plot the above quantity as a function of
$\cos\theta$ for $(\delta\Pi_{BB},\delta\Pi_{WW})=(-0.0029,0)$ and
$(0,-0.0040)$.  (See Table \ref{table:rep}.)
We can see that the angular distribution is affected
differently in two cases.  Thus, a precise study of the angular
distributions provides constraints on 
$\delta\Pi_{BB}$ and $\delta\Pi_{WW}$.

In order to study how well these two parameters are determined, we
perform the following analysis:
\begin{enumerate}
\item We choose several sample points which can explain the diphoton excess.
(See Table  \ref{table:rep}.)
\item
For each sample point, we calculate the new particle contributions to the vacuum polarizations, which we denote by $\overline{\delta\Pi}_{BB}$ and $\overline{\delta\Pi}_{WW}$.

\item 
We estimate the ILC sensitivity for each sample point by using the following quantity:
  \begin{align}
    \Delta \chi^2
    (\delta\Pi_{BB}, \delta\Pi_{WW}; 
    \overline{\delta\Pi}_{BB}, \overline{\delta\Pi}_{WW}) \equiv
    \sum_i
    \frac{(\overline{N}_i^{\rm SM+\psi}-N^{\rm SM+\psi})_i^2}
    {\overline{N}_i^{\rm SM+\psi}+(\epsilon \overline{N}_i^{\rm SM+\psi})^2},
  \end{align}
  where $\overline{N}^{SM+\psi}$ and $N^{SM+\psi}$ are the number of
  $\mu^+\mu^-$ events in each bin evaluated with
  $(\overline{\delta\Pi}_{BB}, \overline{\delta\Pi}_{WW})$ and
  $(\delta\Pi_{BB}, \delta\Pi_{WW})$, respectively.
\end{enumerate}

\begin{table}[t]
  \centering
  \begin{tabular}[h]{l|cccc}
    \hline \hline
    Sample points& 1 & 2  & 3 & 4  \\ \hline \hline
    Representation 
    & $({\bf 1},{\bf 1},1)$ & $({\bf 1},{\bf 3},0)$
    & $({\bf 1},{\bf 1},1)$ & $({\bf 1},{\bf 3},0)$ \\ \hline 
    $m_{\psi}$ [GeV] & 400 & 400 & 600 & 600  \\ \hline
    $N$ & 7 & 3 & 7 & 3 \\ \hline
    $y$ & 0.3 & 0.5 & 0.5 & 1  \\ \hline
    $\Gamma (S\rightarrow\gamma\gamma)$ [MeV] & 1.0 & 0.52 & 0.61 & 0.45  \\ \hline
    $\sqrt{s}$ [GeV] & 500 & 500 & 1000 & 1000  \\ \hline
    $\overline{\delta\Pi}_{BB}(s)$ & $-0.0029$ & 0 & $-0.0066$ & 0  \\ \hline
    $\overline{\delta\Pi}_{WW}(s)$ & 0 & $-0.004$ & 0 & $-0.009$  \\ \hline 
    \hline
  \end{tabular}
  \caption{The parameters of the sample points for our numerical study:
    the representation for ${\rm SU}(3)\times {\rm SU}(2)\times {\rm
    U}(1)_Y$, the fermion mass, the multiplicity $N$, 
    the Yukawa coupling, and 
    $\Gamma (S\rightarrow\gamma\gamma)$.
    We use the sample points 1 and 2 (3 and 4) for the analysis with 
    $\sqrt{s}=500\ {\rm GeV}$ ($1\ {\rm TeV}$).  
    We also show the values of $\delta\Pi_{BB}(s)$ and $\delta\Pi_{WW}(s)$.}
  \label{table:rep}
\end{table}

\begin{figure}[t]
\centering
    \includegraphics[width=18cm]{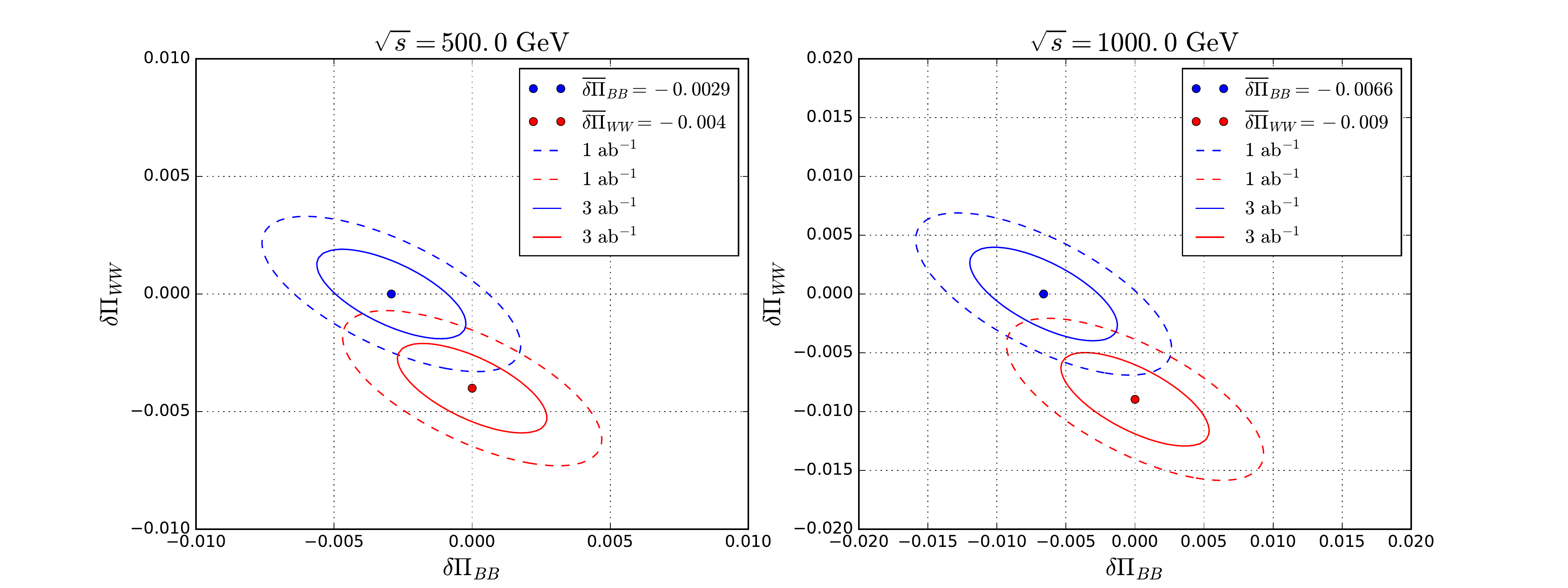}
    \caption{Contours of constant $\Delta \chi^2=5.99$ with the luminosity
  of $1\ {\rm ab}^{-1}$ (dashed) and $3\ {\rm ab}^{-1}$ (solid).
  The blue (red) contours are for Sample points 1 or 3 (2 or 4).
  Here, we take $\epsilon =0$.}
\label{fig:PiBvsPiW}
\end{figure}

In Fig.\ \ref{fig:PiBvsPiW}, the contours of constant $\Delta \chi^2
(\delta\Pi_{BB}, \delta\Pi_{WW}; \overline{\delta\Pi}_{BB},
\overline{\delta\Pi}_{WW})$ are presented on $\delta\Pi_{BB}$ vs.\
$\delta\Pi_{WW}$ plane.  Here, we show $\Delta \chi^2=5.99$, which
gives $95\ \%$ C.L. bounds on the $\delta\Pi_{BB}$ vs.\
$\delta\Pi_{WW}$ plane, taking the luminosity of $1\ {\rm ab}^{-1}$
and $3\ {\rm ab}^{-1}$.  Here, we take $\epsilon =0$ to show the
ultimate sensitivity.  We can see that, with the precision
measurements at the ILC, we will be able to obtain non-trivial
constraint on the $\delta\Pi_{BB}$ vs.\ $\delta\Pi_{WW}$ plane.  In
addition, these results indicate that the ILC may be able to
discriminate models containing new particles with various quantum
numbers.

\section{Summary and discussion}
\label{sec:summary}

In this letter, we have studied the possibility of indirectly probing
the charged particles which are responsible for the diphoton excess
recently reported by the LHC.  If the LHC diphoton excess indicates
the existence of a new resonance $S$ with a mass of $\sim 750\ {\rm
  GeV}$, and also if it has a decay mode
$S\rightarrow\gamma\gamma$, $S$ is likely to couple to new
charged particles whose loop effects induce the coupling between
$S$ and photon.  Even if such charged particles are too heavy to be
accessible with the ILC, they affect the scattering processes
$e^+e^-\to f\bar{f}$ via vacuum polarizations of $\gamma$ and $Z$.
With a precise study of the scattering processes, information about the vacuum
polarization is obtained, from which the existence of the heavy
charged particles can be probed.

We have quantitatively studied such an effect, and shown that the
indirect probe of the charged particles is possible even if they 
are kinematically inaccessible at the ILC.  The effects of the charged
particles on the scattering process is insensitive to the strength of
the interaction between $S$ and the charged particles, but it
depends only on the mass, the multiplicity, and the SU(2)$\times$U(1)$_Y$
representation of the new particles.  We have also shown that the
angular distributions are affected differently by the vacuum
polarizations of SU(2) and U(1)$_Y$ gauge bosons, which makes it 
possible to distinguish signals from new particles with different
quantum numbers.

In our analysis, we have performed our analysis based on LO formulae
of the scattering cross section to demonstrate the expected accuracy
of the indirect probe.  When such an analysis is performed  with real data, however, higher order corrections should be
properly taken into account in order to precisely predict the angular
distribution of the final-state fermions of the scattering processes.
In addition, we have used only the scattering processes with leptonic
final states. 
We may also
be able to use the quark final states taking into account the QCD corrections.

Should the diphoton excess persists with more data at the LHC, 
it is of great importance to probe  the physics behind it.
The precision measurements at the ILC will provide good indirect probes of the 
origin of the diphoton excess, which are complementary to the study at the LHC.


\section*{Acknowledgement}
This work was supported by Grant-in-Aid for Scientific research Nos.\
26104001 (KH), 26104009 (KJB and KH), 26247038 (KH), 26400239 (TM),
26800123 (KH), 16H02189 (KH), and by World Premier International Research Center
Initiative (WPI Initiative), MEXT, Japan.


\end{document}